\documentclass[preprint,showpacs,12pt]{revtex4}
\usepackage{amsmath,amssymb,amsfonts}
\usepackage{graphicx}
\usepackage{hhline}
\usepackage{bm}
        
\usepackage{amsmath}
\usepackage{epsfig}

\begin{document}
\begin{flushleft}
CERN-PH-TH/2012-296\\
\end{flushleft}
\title{Energy reconstruction effects in neutrino oscillation experiments and implications for the analysis}
\vspace*{+0.8cm}
\author {M. Martini}
\affiliation{Institut d'Astronomie et d'Astrophysique, CP-226, Universit\'e Libre de Bruxelles, 1050 Brussels, Belgium}
\author {M. Ericson}
\affiliation{Universit\'e de Lyon, Univ. Lyon 1,
CNRS/IN2P3, IPN Lyon, F-69622 Villeurbanne Cedex, France}
\affiliation{Physics Department, Theory Unit, CERN, CH-1211 Geneva, Switzerland}
\author {G. Chanfray}
\affiliation{Universit\'e de Lyon, Univ. Lyon 1,
CNRS/IN2P3, IPN Lyon, F-69622 Villeurbanne Cedex, France}

\begin{abstract}
Data on neutrino oscillation often involve reconstructed neutrino energies while the analysis implies 
the real neutrino energy. The corrections corresponding to the transformation from real to reconstructed energy are discussed in the case of Cherenkov 
detectors where  multinucleon events appear as quasielastic ones. These corrections  show up as a tendency for the events to escape the region of high flux, with a clear preference for the low energy side. This is an effect of the multinucleon component of the quasielastic cross section. We have applied our corrections to 
the T2K and MiniBooNE data for electron appearance or $\nu_\mu$ disappearance data. 
We show that the inclusion of this correction in the analysis is expected to lead to an increase of the best fit oscillation mass parameters, particularly pronounced 
for the MiniBooNE neutrino data. This inclusion in the analysis of the MiniBooNE neutrino data should improve the compatibility with the existing constraints.
\end{abstract}

\maketitle

\section{Introduction}
The interpretation of neutrino oscillation experiments requires the determination of the neutrino energy which enters the expression of the oscillation length. 
This determination is often done through ``quasielastic events''  and the reconstructed energy hypothesis in which the neutrino interaction in the 
nuclear target with lepton production is supposed to take place on a nucleon at rest. In this case the only measurement of the lepton variables, 
its energy and emission angle, allows the reconstruction of the  neutrino energy. 
In Cherenkov detectors  ``quasielastic events'' are defined as those in which the emission product only include one lepton, the ejected nucleons 
being unobservable for these types of detectors. In this case the definition of quasielastic events incorporates multinucleon ones 
which are indistinguishable from the genuine quasielastic ones. 
We have suggested  \cite{Martini:2009uj} that the MiniBooNE axial mass anomaly \cite{AguilarArevalo:2010zc} 
in the quasielastic cross section could be accounted for by the inclusion of the multinucleon channel
and we have been able to reproduce \cite{Martini:2009uj,Martini:2011wp} the total 
``quasielastic'' cross section as well as the measured double differential cross section 
without any modification of the axial mass.
In a more recent work \cite{Martini:2012fa} we have addressed the question of the energy reconstruction, taking into account  the fact that not 
all events being real quasielastic ones,  the usual reconstruction method becomes questionable. 
We have explored,  for charged current neutrino reactions, the corrections to this method 
in the form of a probability distribution, $F(E_\nu, \overline{E_\nu})$, to have a real energy $E_\nu$  starting from a  reconstructed value $\overline{E_\nu}$. 
We have shown that this distribution can be expressed in terms of the double differential neutrino nucleus cross section with respect to the energy $\omega$ transferred to the nucleus and the lepton emission angle $\theta$,
$\frac{d^2 \sigma}{d \omega  ~d\mathrm {cos}\theta}$. The distribution also involves the neutrino flux distribution $\Phi(E_{\nu})$. 
Similar approaches taking as well into account the multinucleon contribution have followed \cite{Lalakulich:2012ac,Nieves:2012yz}. 

 The double differential cross section  $\frac{d^2 \sigma} {d E_{\mu } ~d\mathrm {cos}\theta}$ for muon production has been measured by MiniBooNE \cite{AguilarArevalo:2010zc}. 
Although the ponderations by the neutrino flux do not allow a direct insertion of this experimental data to derive the energy distributions, 
the fact that our theoretical model is able to account in a satisfactory way for the MiniBooNE data of this double differential cross section \cite{Martini:2011wp} allows a certain degree of confidence in the distributions that we have obtained. The most spectacular influence of the multinucleon component in the distribution is the existence of a tail in the neutrino energy region above the reconstructed energy. 
With the evaluation of these probability  distributions we could transform the  experimental distributions such as those given by 
MiniBooNE \cite{AguilarArevalo:2007it,AguilarArevalo:2008rc,AguilarArevalo:2012va} or T2K \cite{Abe:2011sj,Abe:2012gx}, 
in terms of reconstructed neutrino energies, to smeared ones expressed in terms of the true energies which may then be confronted to oscillation models. 
  We have applied \cite{Martini:2012fa} 
the smearing procedure  to the oscillation electron events issued from the oscillation of a  muon neutrino beam in the case of  
 the T2K experiment  where  electrons from the interaction  of electron neutrinos  attributed to an oscillation  process 
of a muon neutrino beam have been observed and their distribution in terms of the reconstructed energy given \cite{Abe:2011sj}. 
Here the disappearance effect is well established and the oscillation parameters known to a good accuracy. 
Hence the electron neutrino energy distribution which enters the smearing function is predictable from the muon neutrino one.  We have applied our smearing procedure to extract the distribution in terms of the real neutrino energy. After this procedure  the shape of the distribution could be successfully compared with the expected theoretical predicted shape, product of our total quasielastic cross section with the electron neutrino distribution generated by the oscillation of muon neutrinos. This consistency confirms the validity of the assumptions made: the observed electron events do come from an oscillation phenomenon and  the oscillation mass parameter has the right magnitude.
 The second case is the MiniBooNE one where excess electron events  attributed to the oscillation of the muon beam have also been observed 
\cite{AguilarArevalo:2007it,AguilarArevalo:2008rc}. 
Here one deals with a short baseline experiments, $L = 541$ m,  and  the grounds for describing these oscillations are not as established. 
Short baseline oscillation phenomena,  if they exist, involve sterile neutrinos with large values of the $\Delta m^2$, in the eV$^2$ range. 
But this value, hence the electron neutrinos energy distribution, is uncertain. We remind that our smearing procedure depends on this distribution. 
Assuming as an example that it followed the muon neutrino one we have applied the smearing procedure to the experimental MiniBooNE distribution. 
The data present a low energy anomaly with an excess of low reconstructed energy electron events. After our smearing procedure the excess of low energy electron neutrino events which raises a problem in the oscillation models, is  pushed towards higher energy, making it possibly compatible with oscillation models. Of course this conclusion was not reliable as the assumption made on the electron neutrino spectrum was not based on oscillation models and therefore totally arbitrary. Indeed in this case we observe that, contrary to the T2K case, there is no consistency at all between the smeared experimental distribution and the theoretical one, product of the total  charged current ``quasielastic'' cross section for electron neutrino by their energy distribution, $\sigma _{\nu_e}(E_{\nu_e}) \Phi(E_{\nu_e})$, the theoretical  one extending much further in energy. This is not a surprise since the flux  model was totally arbitrary. However this study indicated the possibility of improving the description of the anomaly through the reconstructed energy corrections.
In the present work we discuss in a more realistic way the influence of the smearing procedure on the two experiments both for the muon events from the $\nu_\mu$ beams and for the electron ones coming from the oscillation. The  aim is to  show the possible influence of  nuclear physics effects on the determination of the oscillation parameter.
 For MiniBooNE we discuss the cases both neutrinos and antineutrinos. In the first one we have the information that our description of the neutrino nucleus interaction is able to reproduce 
the double differential cross section with respect to the muon energy and emission angle. In the case of  antineutrinos no such cross check can be made. 
The total "quasielastic" cross section is not published nor is the double differential cross section. 
We have to rely on a purely theoretical approach, which was described in our Refs. \cite{Martini:2009uj,Martini:2010ex}. 
Its main features are summarized in the following section.
\section{Theory}
 
In the following  the word neutrino will mean neutrino or antineutrino, the lepton can be either a $\mu^{-,+}$ or an electron (positron), 
depending on the reaction in question. 
For MiniBooNE the oscillation events concern electron neutrinos and the interesting events represent the production of electron arising from 
the interaction of electron neutrinos in the Cherenkov detector.
 The neutrino cross sections on nuclear targets are expressed in terms of nuclear responses, which represent the inelastic cross sections for a given type of coupling of the probe. For the charged current interaction case the probes have a purely isovector character and they can involve or not the nucleonic spins. Our description of Refs. \cite{Martini:2009uj,Martini:2010ex,Martini:2011wp} introduces the multinucleon component only in the spin isospin response.  For the comparison between neutrinos and antineutrinos the important point  is the role played by this isovector response relative to that of the spin isospin ones. Due to the negative sign of the axial vector interference term for antineutrinos the spin isospin contribution is diminished in the last case with respect to the isovector component. Consequently in our description the relative role of the multinucleon part is smaller for antineutrinos. 
This particular point of our treatment of the multinucleon channel has been challenged \cite{Nieves:2011pp,Amaro:2011aa}. 
Experimental data will select between different approaches but, whatever the outcome, the method that we introduce for extracting a real distribution energy distribution from the reconstructed one and our general conclusions on the reconstruction effect remain valid. The difference of nuclear effects between neutrinos and antineutrinos is a very important point as it can mimic or mask the display of CP violating effect which introduces an asymmetry between $\nu$ and $\bar{\nu}$. 

 In our last work \cite{Martini:2012fa} we started from the electron events experimental distributions in terms of the reconstructed energy. We transformed these into  distributions in terms of the real neutrino energy. We remind that this transformation itself depends on the electron neutrino energy distribution hence on the oscillation parameter. Then we evaluated the theoretical expectations of the distribution with the same oscillation parameter. We could then confront theory and experiment.
Although perfectly valid this method has the following drawback. When data evolve, the whole smearing procedure has to be redone. Moreover as each reconstructed energy bin influences the whole real energy distribution, negative values in certain bins influence the whole distribution. 
Here we will proceed differently. 
Indeed, as we observed in \cite{Martini:2012fa}, the procedure is completely reversible and can be used in both directions.
Here we calculate the theoretical prediction for electron events energy distribution for a given value of the oscillation parameter. 
We then transform this distribution into one in terms of the reconstructed energy value, which can be directly compared to the experimental distribution. In principle we are then in a situation to investigate which oscillation parameter best fits the data.  
We also apply our smearing procedure to disappearance effects  for the muon neutrinos in the T2K beam.

\subsection{Formalism}

\begin{figure}
\begin{center}
\includegraphics[width=12cm,height=8cm]{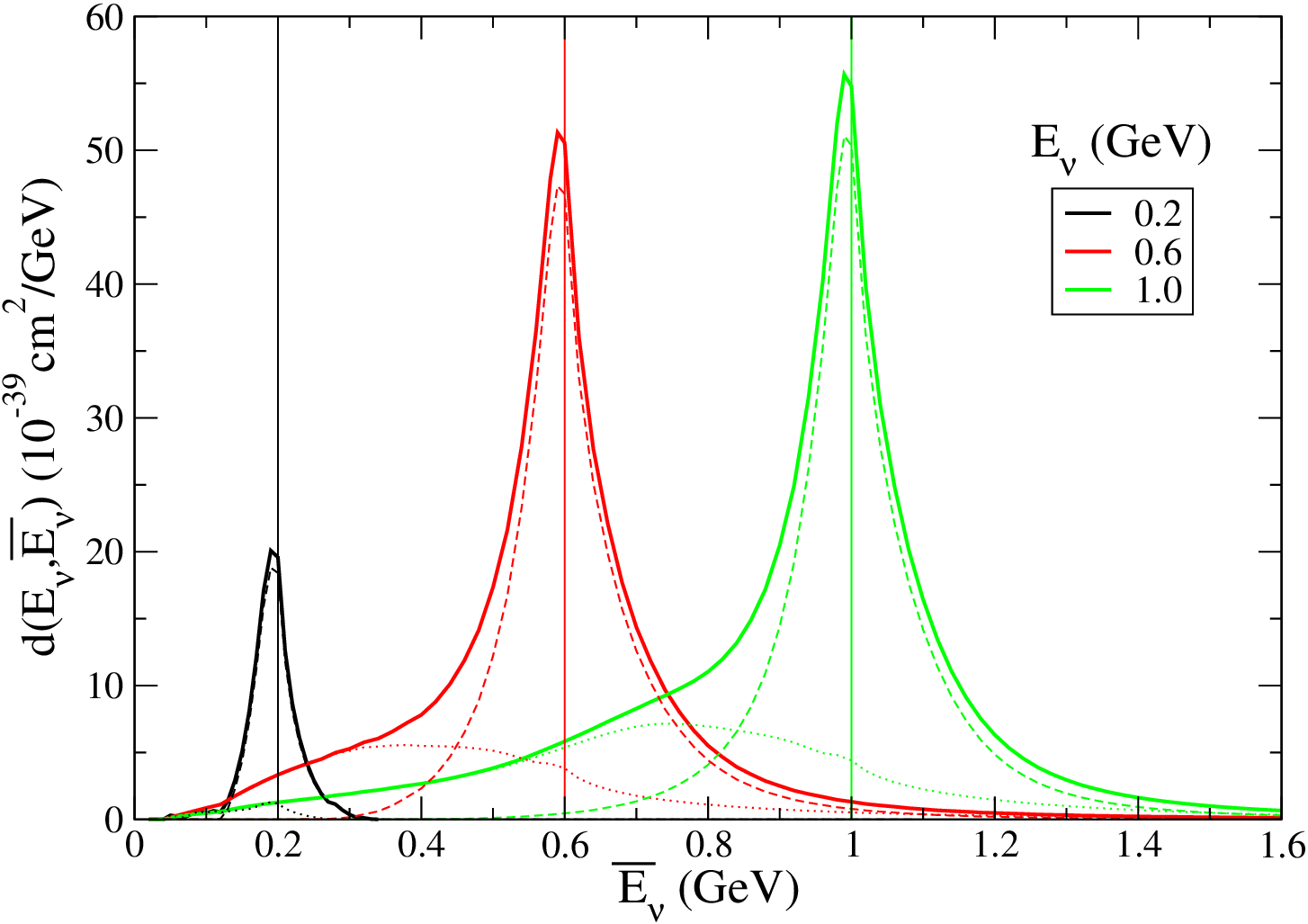}
\caption{(Color online) The spreading function $d(E_{\nu},\overline {E_{\nu}})$ of Eq. (\ref{rho_enubar}) 
per neutron of $^{12}$C in the case of electrons evaluated for three $E_{\nu}$ values. 
The genuine quasielastic (dashed lines) and the multinucleon (dotted lines) contributions are also shown separately.
}
\label{fig_integral_vs_erec}
\end{center}
\end{figure}

 The number of charged current events in a target for neutrinos of energy  between $E_\nu$  and $E_\nu+d E_{\nu }$, 
for an energy transferred to the nuclear system, $\omega$, and a lepton emission angle  $\theta$, is related to the double differential cross section by :
\begin{equation}
 g(E_{\nu}, \omega, \cos \theta)~d E_{\nu }~ d \omega~ d\mathrm{cos} \theta  = 
\frac{d^2 \sigma}{d \omega  ~d\mathrm{cos}\theta}~ \Phi(E_{\nu})~ d E_{\nu }~ d \omega~ d\mathrm{cos} \theta.
\end{equation}
 The quantity $g$ is the triple density, in terms of the three variables, $E_\nu$, $\omega $ and $\cos \theta$. 
For our problem  it is convenient to switch to another set of variables, $E_\nu, E_l $(the energy of the lepton produced) and the reconstructed neutrino energy $\overline E_{\nu}$.
The relations between the two set of variables are  firstly :   $\omega =E_{\nu }- E_l $. In addition $\cos \theta $ is related to   the new variables $E_l $ and   $\overline {E_{\nu}}$ by:
\begin {equation}
\label{eq_erec}
\overline {E_{\nu}} P_l \cos \theta + M (\overline E_{\nu}-E_l)-\overline E_{\nu}E_l +  \frac{{m_l}^2}{2}=0,
\end {equation} 
where $P_l $ is the lepton momentum, $m_l$ the charged lepton mass and $M$ the nucleon mass. 
The modulus of the Jacobian for these variables transformation is   
$(ME_l - m_l^2/2) ( \overline E_\nu^2 P_l)^{-1}$ , and the new density $G ( E_{\nu}, E_l , \overline E_{\nu}) $ is 
\begin {equation}
G( E_{\nu}, E_l , \overline E_{\nu})~ d E_{\nu}~ dE_l  ~ d \overline E_{\nu} =
 d E_{\nu}~ dE_l ~  d \overline E_{\nu} ~ \Phi (E_{\nu})~ \frac{ME_l - m_l^2/2}
  {\overline E_{\nu}^2 P_l} \left[\frac{d^2 \sigma}{d \omega  ~d\cos \theta }\right]_{\omega 
  = E_{\nu}-E_l,~\mathrm{cos}\theta=\cos \theta(E_l, \overline E_\nu)}, 
\end {equation} 
where $\cos \theta(E_l, \overline E_\nu)$ is the solution of Eq. (\ref{eq_erec}).
After integration over the lepton energy this density can be used in both directions. 
Either to extract a distribution in terms of the real neutrino energy  from a distribution in reconstructed energies, 
as was done in our previous work \cite{Martini:2012fa} where we had used normalized probabilities. 
Or, in the opposite direction, we start from a theoretical distribution expressed with real energies 
then we perform the smearing procedure to deduce the corresponding distribution of the events in terms of the reconstructed energy. For this we integrate 
  over the lepton energy and over the real  neutrino energy distribution, which 
   provides the distribution,  $D_{rec}(\overline {E_{\nu} })$, in terms of the reconstructed energy which can  be compared to the data 
   \begin{equation}
  \label{rho_enubar}
 {D_{rec}(\overline {E_{\nu} })} =   \int  
 d E_{\nu}   \Phi (E_{\nu})  \int _{E_l^{min}}^{E_l^{max}} d E_l\frac{ME_l - m_l^2/2}{ \overline E_{\nu}^2 P_l}\left[\frac{d^2 \sigma}{d \omega  ~d\cos \theta }\right]_{\omega=E_{\nu}-E_l,~\mathrm{cos}\theta=\cos \theta(E_l, \overline {E_\nu })}, 
\end {equation} 
where the quantities $E_l^{min}$ and $E_l^{max}$ are the minimum and maximum values of the charged lepton energy for a given value 
of $ \overline {E_{\nu} }$. They  are obtained by taking 
$\cos\theta= 1(-1)$ in the Eq. (\ref{eq_erec}),  with the  additional restriction, $m_l <E_l < E_\nu$. 
The second integral on the r.h.s. of Eq. (\ref{rho_enubar}), which represents the spreading function, depends on $E_{\nu}$ and  $ \overline {E_{\nu} }$; 
we denote it as $d(E_{\nu},\overline {E_{\nu}})$. 
We give  in Fig. \ref{fig_integral_vs_erec} some examples of its 
 $ \overline {E_{\nu} }$ 
 dependence for several $ E_\nu$ values. 
The np-nh low energy tail  is the counterpart, in these variables, of the high  energy one that we stressed in our previous work \cite{Martini:2012fa}.

 \section{Applications}
 \subsection{T2K}
\begin{figure}
\begin{center}
\includegraphics[width=12cm,height=8cm]{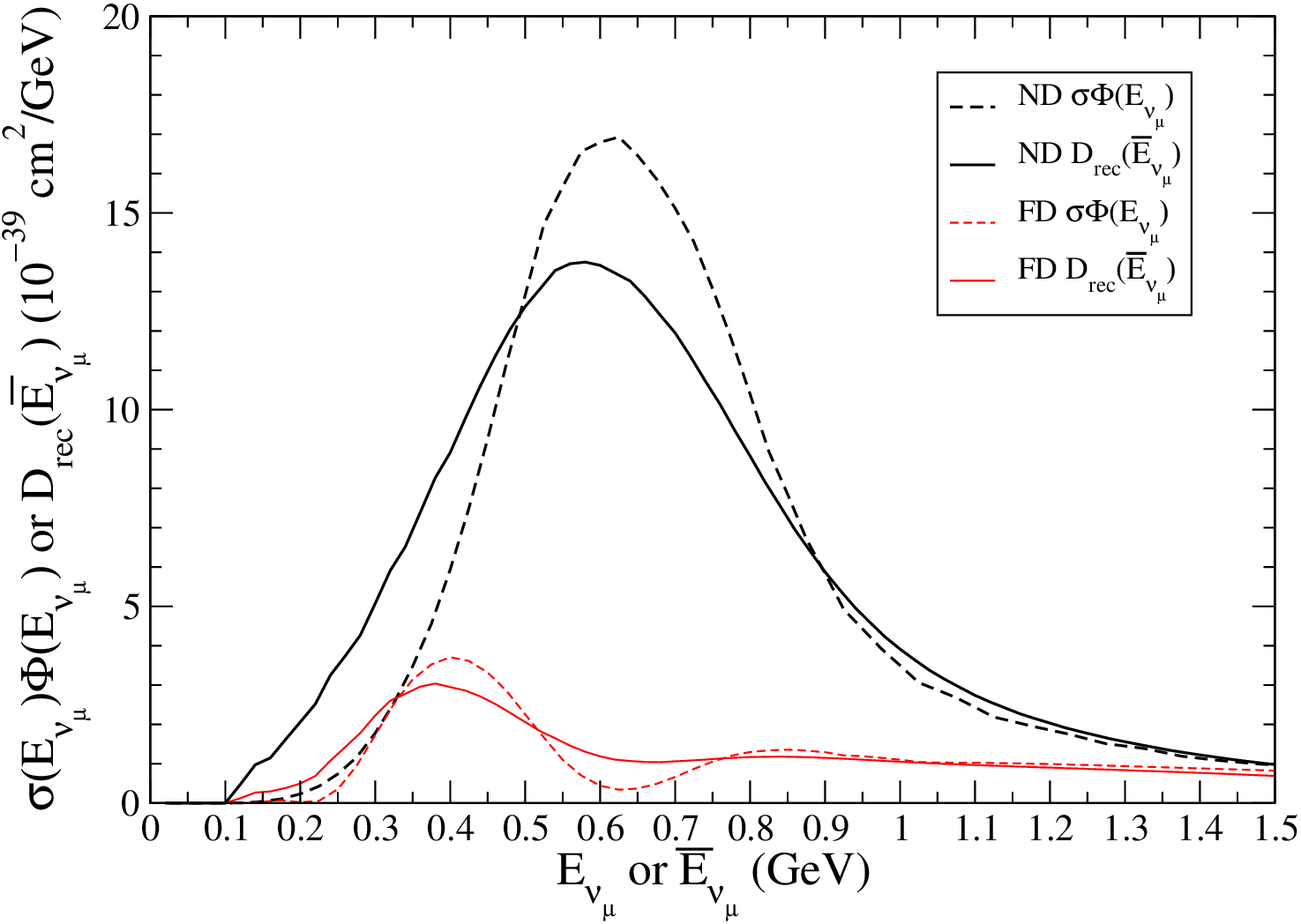}
\caption{(Color online) T2K distributions per neutrons of muon events before (dashed lines) and after (continuous lines) 
reconstruction in the near and far detector, evaluated with the parameters of the T2K best fit \cite{Abe:2012gx}.
}
\label{fig_t2k_sigma_flux_3_cases}
\end{center}
\end{figure}

 Here the situation is relatively simple as one deals with a long baseline experiment \cite{Abe:2012gx,Abe:2011sj} with oscillation mass parameters already known to a  good accuracy. 
We have pointed out  \cite{Martini:2012fa} the interest of the study for T2K of the muon events spectrum both in the close detector and in the far detector since the two corresponding muonic neutrino
 beams have  different energy distributions.  The study of the reconstruction influence on the electron events in the far SuperKamiokande detector was performed in our Ref. \cite{Martini:2012fa}, it is discussed again here in our new reversed perspective. The two muon beams in the close and far detectors and the oscillated electron beam at the far detector  having widely different energy distributions, the effect of the reconstruction is expected to differ in all three.
\begin{figure}
\begin{center}
\includegraphics[width=12cm,height=8cm]{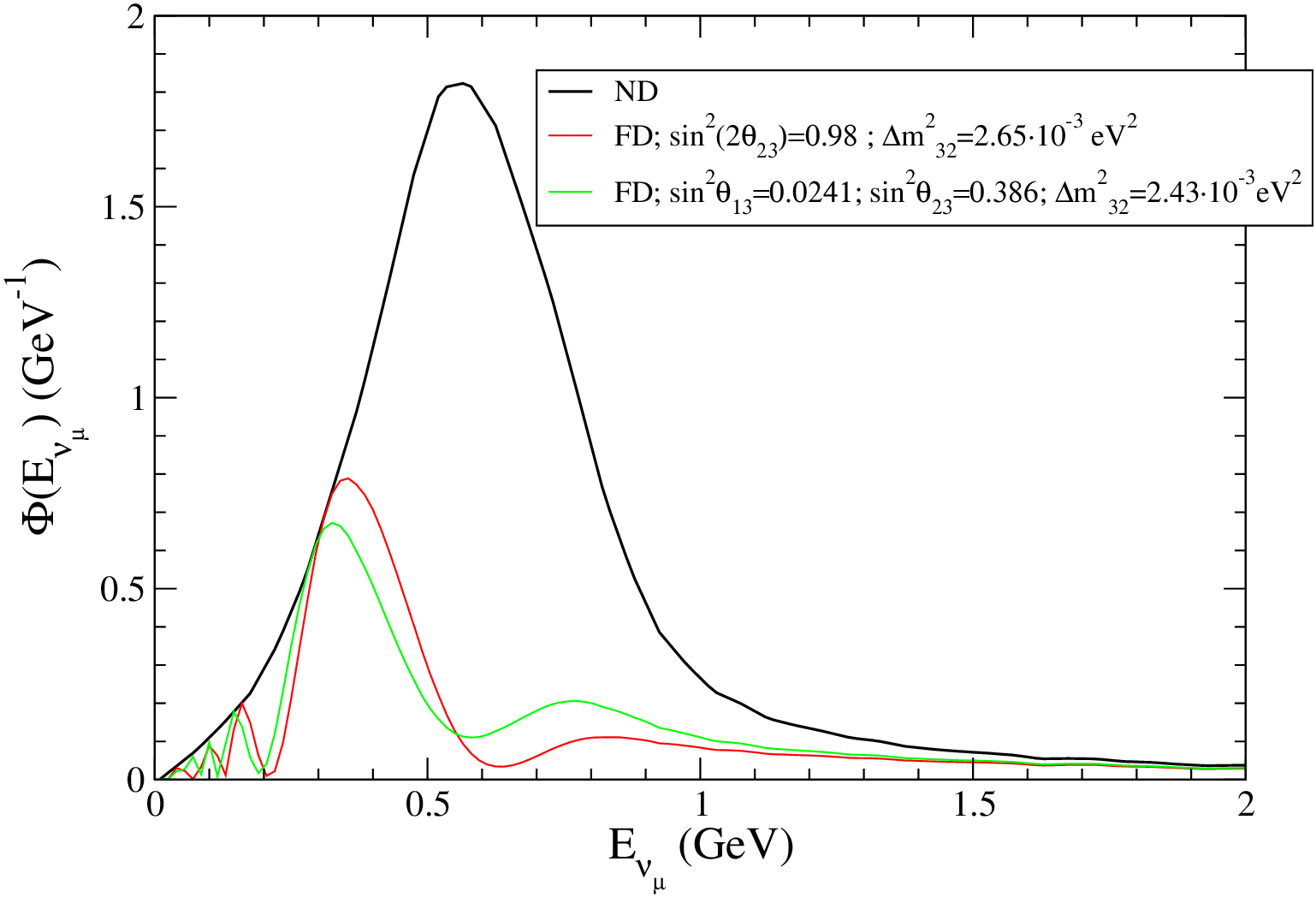}
\caption{(Color online) T2K $\nu_\mu$ flux energy distributions in the near detector (ND) and far detector (FD) for tho sets 
of oscillation parameters according to Ref. \cite{Abe:2012gx} and Ref. \cite{Fogli:2012ua}.
}
\label{fig_t2k_flux_numudisapp}
\end{center}
\end{figure}
 The muon neutrino energy distribution in the close detector, normalized with an energy integrated value of unity, 
$\Phi_{\nu_\mu}(E_{\nu_\mu})$ is represented in 
Fig. \ref{fig_t2k_flux_numudisapp} as a function of $E_{\nu_\mu}$. 
At the arrival in the far detector it is reduced by a large factor which depends on the oscillation parameters and its expression is :
\begin{equation}
\label{eq_t2k_sk_flux_mu}
\Phi^{FD}_{\nu_\mu}(E_{\nu_\mu})=\left[1-4 \cos^2 \theta_{13} \sin^2\theta_{23} (1-\cos^2 \theta_{13} \sin^2\theta_{23}) 
\sin^2\left(\frac{\Delta m_{32}^2 L}{4 E_{\nu_\mu}}\right)\right]\Phi^{ND}_{\nu_\mu}(E_{\nu_\mu}).
\end{equation} 
We use this expression as a definition of the far detector flux.                                  
We have kept in this expression  the influence of the angle $\theta_{13}$, which is now measured \cite{An:2012eh}: $\sin^2 \theta_{13}=0.024 \pm 0.004$.  
Its effect is not totally negligible and it partly fills the dip of the energy distribution in the far detector.
 The oscillated $\nu_\mu$ distribution is  shown as well in Fig. \ref{fig_t2k_flux_numudisapp} for the values of the parameters 
of Ref. \cite{Fogli:2012ua}    
 and also for the best fit values of T2K \cite{Abe:2012gx} where the effect of $\theta_{13}$ is ignored. 
The products  $\sigma (E_{\nu_\mu}) \Phi_{\nu_\mu}(E_{\nu_\mu})$ which represent  the distributions of muon events before reconstruction in the close and far detector are shown in Fig. \ref{fig_t2k_sigma_flux_3_cases}.    
%
  We now apply  our smearing procedure to these distributions and we obtain the smeared curves also shown in Fig. \ref{fig_t2k_sigma_flux_3_cases}. 
The salient features are the broadening effects. 
In the close detector there is  clear low energy enhancement,  an effect of the multinucleon component. 
In the far detector, where the unsmeared distribution displays  a pronounced dip, the smeared one acquires a low energy tail and the middle hole is largely filled,  an effect  also largely due to the np-nh  cross section.  
   All these smearing effects can be described as a tendency to escape the regions of high fluxes when one goes from true to reconstructed energies. We remind that the opposite transformation from reconstructed to true energy tends instead to concentrate the events in the regions of high flux \cite{Martini:2012fa}. 
\begin{figure}
\begin{center}
\includegraphics[width=12cm,height=8cm]{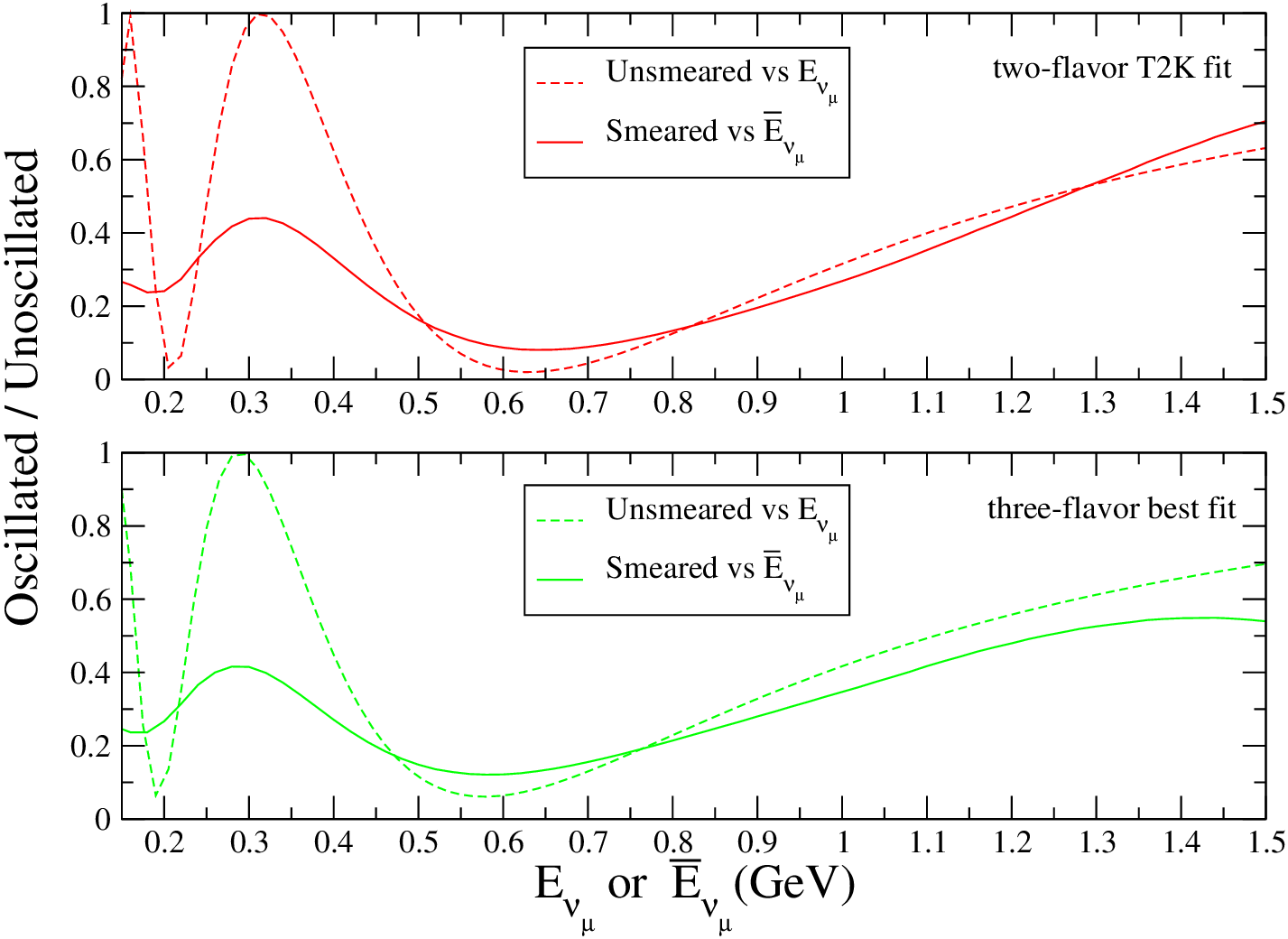}
\caption{(Color online) $\nu_\mu$ disappearance oscillation probability with (continuous line) and without (dashed line) smearing.
}
\label{fig_art_t2k_ratio}
\end{center}
\end{figure}
 The ratio of the smeared distributions in the far and near detector displayed in Fig. \ref{fig_art_t2k_ratio} is far from the unsmeared one, i.e., 
from the oscillation disappearance  factor.
\begin{figure}
\begin{center}
\includegraphics[width=12cm,height=8cm]{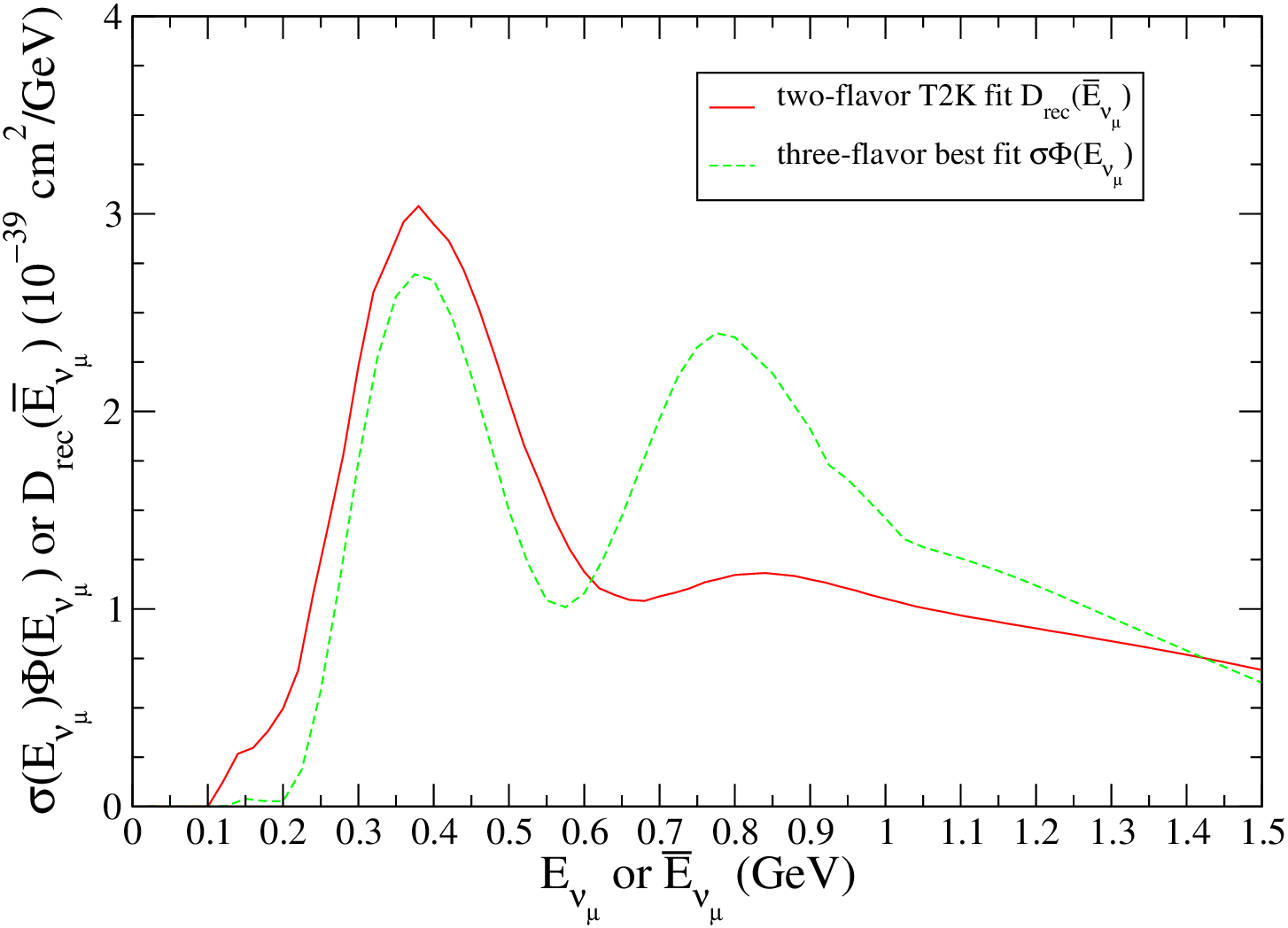}
\caption{(Color online) The smeared distribution, ${D_{rec}(\overline {E_{\nu} })}$, with the T2K two-flavor best fit \cite{Abe:2012gx} 
and the unsmeared curve, $\sigma (E_{\nu_\mu}) \Phi_{\nu_\mu}(E_{\nu_\mu})$, with the  best fit value  of Ref. \cite{Fogli:2012ua}. 
The respective mass values are $\Delta m^2_{32}=2.65 ~10^{-3}$ eV$^2$ and $\Delta m^2_{32}=2.43 ~10^{-3}$ eV$^2$.
}
\label{fig_art_t2k_2fsm_3funsm}
\end{center}
\end{figure}

\begin{figure}
\begin{center}
\includegraphics[width=12cm,height=8cm]{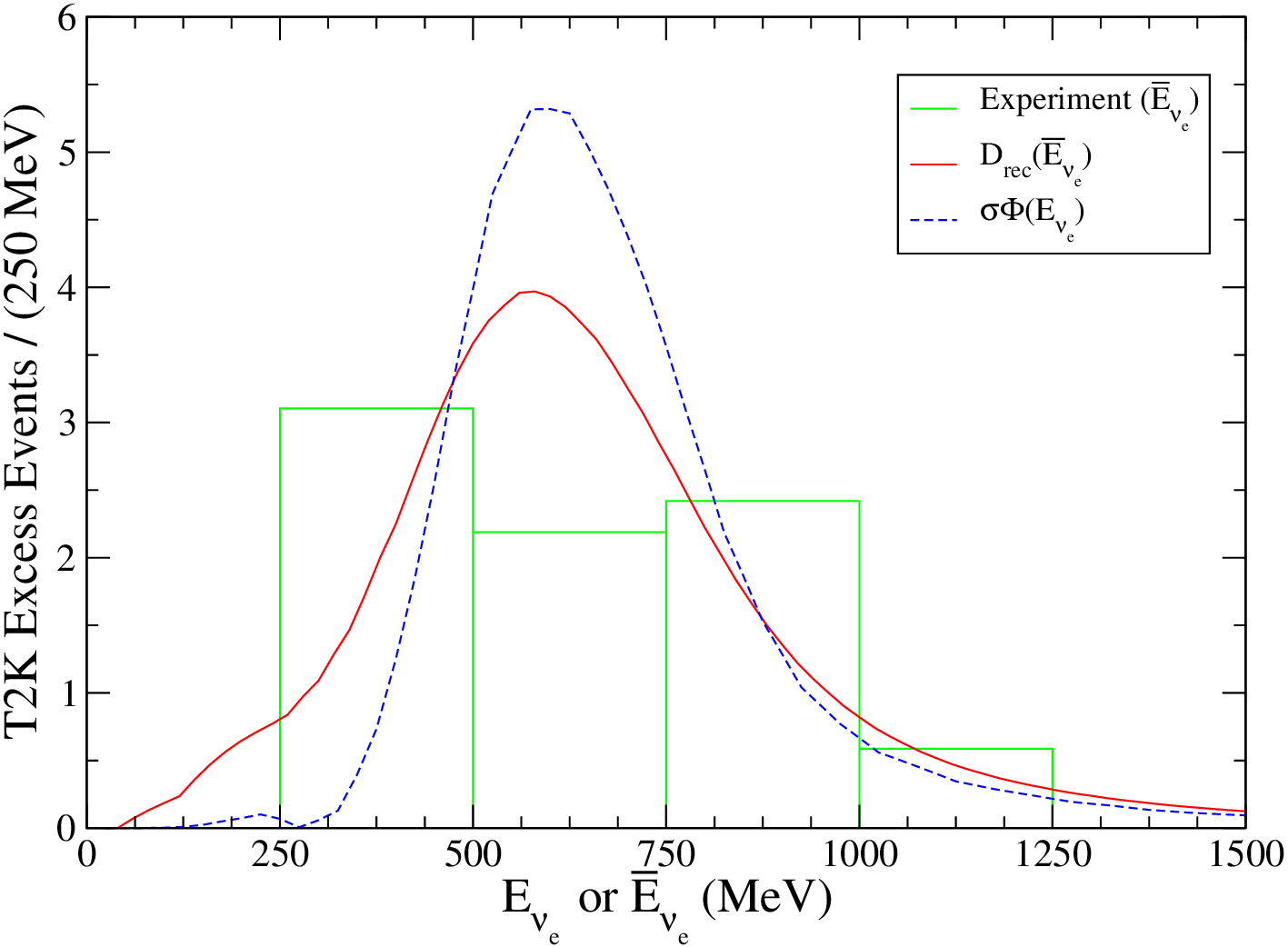}
\caption{(Color online) T2K oscillation electron events energy distributions before (dashed lines) and after (continuous lines) smearing. 
The experimental histogram is also shown.
}
\label{fig_t2k_nue}
\end{center}
\end{figure}

 Notice that the displacement in the far detector of the low energy peak towards smaller values by the smearing can be, to some extent, simulated 
 by a decrease of the mass value in the unsmeared situation. As an illustration we compare in Fig. \ref{fig_art_t2k_2fsm_3funsm} the smeared curve with the T2K best fit, which has a mass value 
$\Delta m^2_{32}=2.65~10^{-3}$ eV$^2$ and the unsmeared curve with the best fit value  $\Delta m^2_{32}=2.43~ 10^{-3}$ eV$^2$ of Ref. \cite{Fogli:2012ua}. 
The equivalence only holds in the first peak region but it is likely to that the inclusion of the reconstruction effects the analysis of the T2K disappearance data will result in a slight increase of the mass value. This investigation will be the object of a future work, following the approach of \cite{FernandezMartinez:2010dm,Meloni:2012fq}. We will see that in the MiniBooNE case the equivalence between the introduction of the reconstruction effect and a lowering of the mass value in the unsmeared situation can be total.  
Turning now to the electron events distribution we display it in Fig. \ref{fig_t2k_nue} 
without and with the reconstruction correction. We also show the actual experimental histogram \cite{SakashitaICHEP2012} 
of significant events (i.e. with the background subtraction). The theoretical curve has been normalized to the same total number of these events. 
Again here the reconstruction correction tends to make events leak outside the high flux region, in agreement with the observed trend.

 \subsection{MiniBooNE}
 
 We turn to the MiniBooNE data which are now available for neutrino and antineutrinos \cite{AguilarArevalo:2012va}. 
In both cases an electron excess has been observed and interpreted in terms of oscillations induced by the existence of sterile neutrinos. In the $3+1$ hypothesis of only one sterile neutrino the electron neutrino flux is given by an expression similar to the usual two-flavor expression for active neutrinos. 
Starting from a muon neutrino energy distribution  $\Phi_{\nu_\mu}(E_{\nu_\mu})$, the electron one generated by the oscillations is given by
 the following expression:
\begin{equation}
\label{flux_osc_41}
\Phi_{\nu_e}(E_{\nu_e})= \Phi_{\nu_\mu}(E_{ \nu_{\mu}})\sin^2(2\theta_{41}) \sin^2\left(\frac{\Delta m_{41}^2 L}{4 E_\nu}\right), 
\end{equation}
 where $\Delta m_{41}^2$ is the difference of the square mass of the sterile neutrino and  $m_1^2$ (or  $m_2^2$).
    Short baseline oscillation experimental results imply large values of  $\Delta m_{41}^2$, in the eV$^2$ range. 
Constraints on this parameter have been provided by a series of data (see for example Ref. \cite{Abazajian:2012ys} for a review or Ref. \cite{Giunti:2011bx})
and more recently by the ICARUS experiment \cite{Antonello:2012pq}. Here our aim is to explore the oscillation phenomenon taking into account the problems of energy reconstruction, which has not been done previously. 
The data  give the distribution of electron events as a function of the reconstructed neutrino energy. 
They have revealed a striking feature, denoted as the MiniBooNE anomaly \cite{AguilarArevalo:2007it,AguilarArevalo:2008rc}, namely an excess of events at low energies. 
There is instead a shortage of events above  $ \overline {E_{\nu_e} }\gtrsim  450$ MeV.  This has been the object of an intense debate. The low energy behavior of the data favors small values of the mass parameter, $\Delta m_{41}^2 \simeq 0.1$ eV$^2$,  
which concentrate the $\nu_e$ flux from the oscillation at low energies. But small values imply, in order to have enough events, large values of the $\sin^2(2\theta_{41})$ which are not compatible with the constraints from other sets of data \cite{Abazajian:2012ys}. 
\begin{figure}
\begin{center}
\includegraphics[width=12cm,height=8cm]{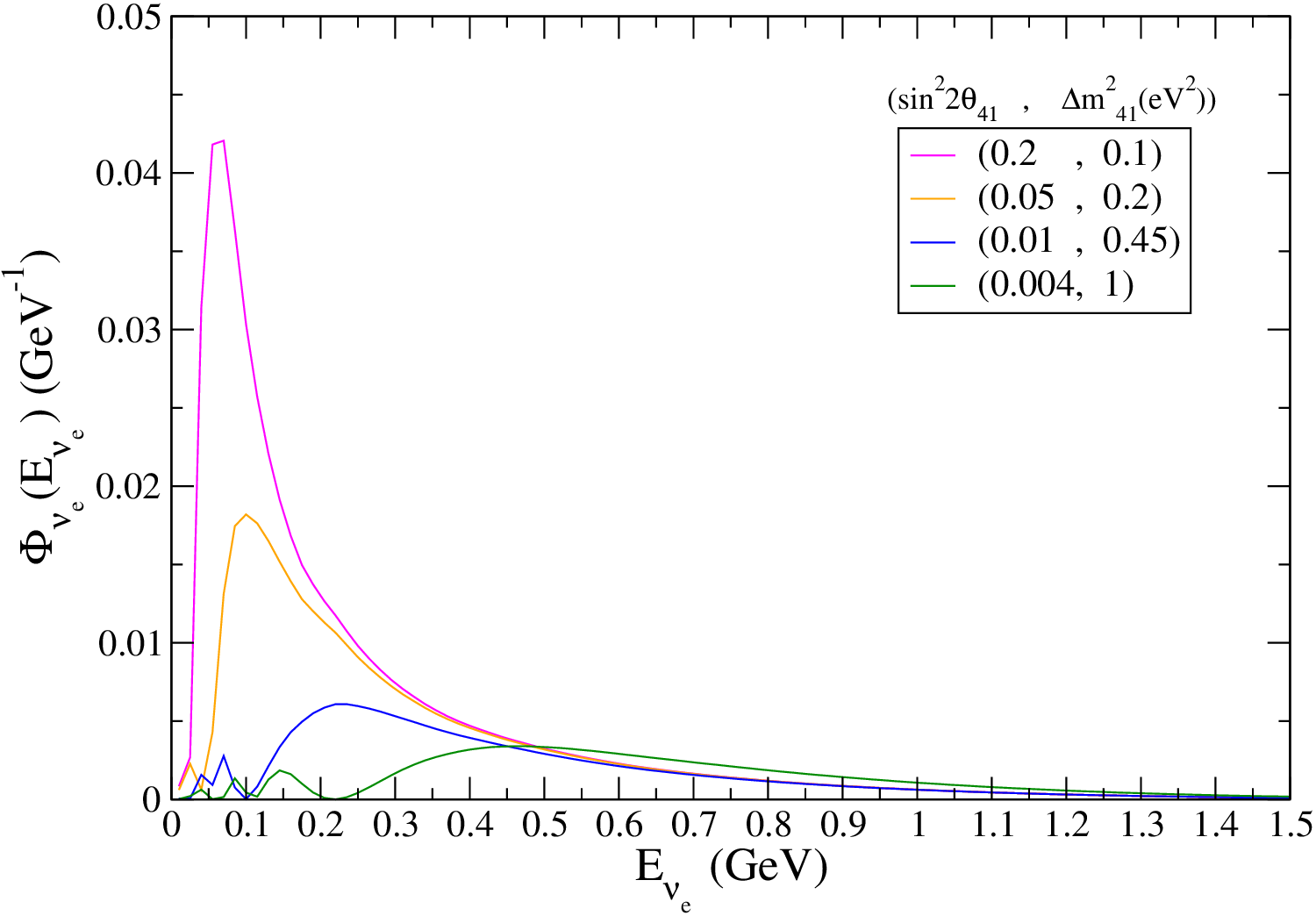}
\caption{(Color online) Electron neutrino energy distributions as a function of $E_{\nu_e}$ from the oscillation expression of 
Eq. (\ref{flux_osc_41}) for four set of values of the oscillation parameters.
}
\label{fig_art_minib_flux_osc}
\end{center}
\end{figure}
  Figure \ref{fig_art_minib_flux_osc}  illustrates the electron neutrino beam energy distribution as a function of the energy obtained from the oscillation expression of Eq. (\ref{flux_osc_41}) for some set values of  $\sin^2(2\theta_{41})$ and $ \Delta m_{41}^2$. Two values 
  of the chosen set are reference  values for LSND.
We have used as the input of Eq. (\ref{flux_osc_41}) the MiniBooNE muon flux 
$\Phi_{\nu_\mu}(E_{ \nu_{\mu}})$, as given in \cite{AguilarArevalo:2010zc}, normalized with an energy integrated value of unity.    
 For large values  of the $ \Delta m_{41}^2$ parameter, the corresponding $\sin^2$ term in the expression of the flux 
can take zero values in the relevant neutrino energy range, 0.2 GeV $<E_{\nu_e}<$ 1 GeV. 
For  small values instead, it is a regularly decreasing function of the energy in the energy region above 0.2 GeV.  
When $ \Delta m_{41}^2$ reaches values smaller than  $\simeq$ 0.1 eV$^2$ the $\sin$ function can be expanded 
and the energy distribution goes as $1/E_{\nu_e}^2$. 
Once this situation is attained, for the concentration of the events at low energy there is no gain in lowering further the value of the mass, 
the energy distribution of the beam remaining the same, and it is possible to adjust the associated value of $\sin^2(2\theta_{41})$ 
to reach the required magnitude. We remind that the smearing effect does not depend on the value of $\sin^2(2\theta_{41})$.

\begin{figure}
\begin{center}
\includegraphics[width=12cm,height=8cm]{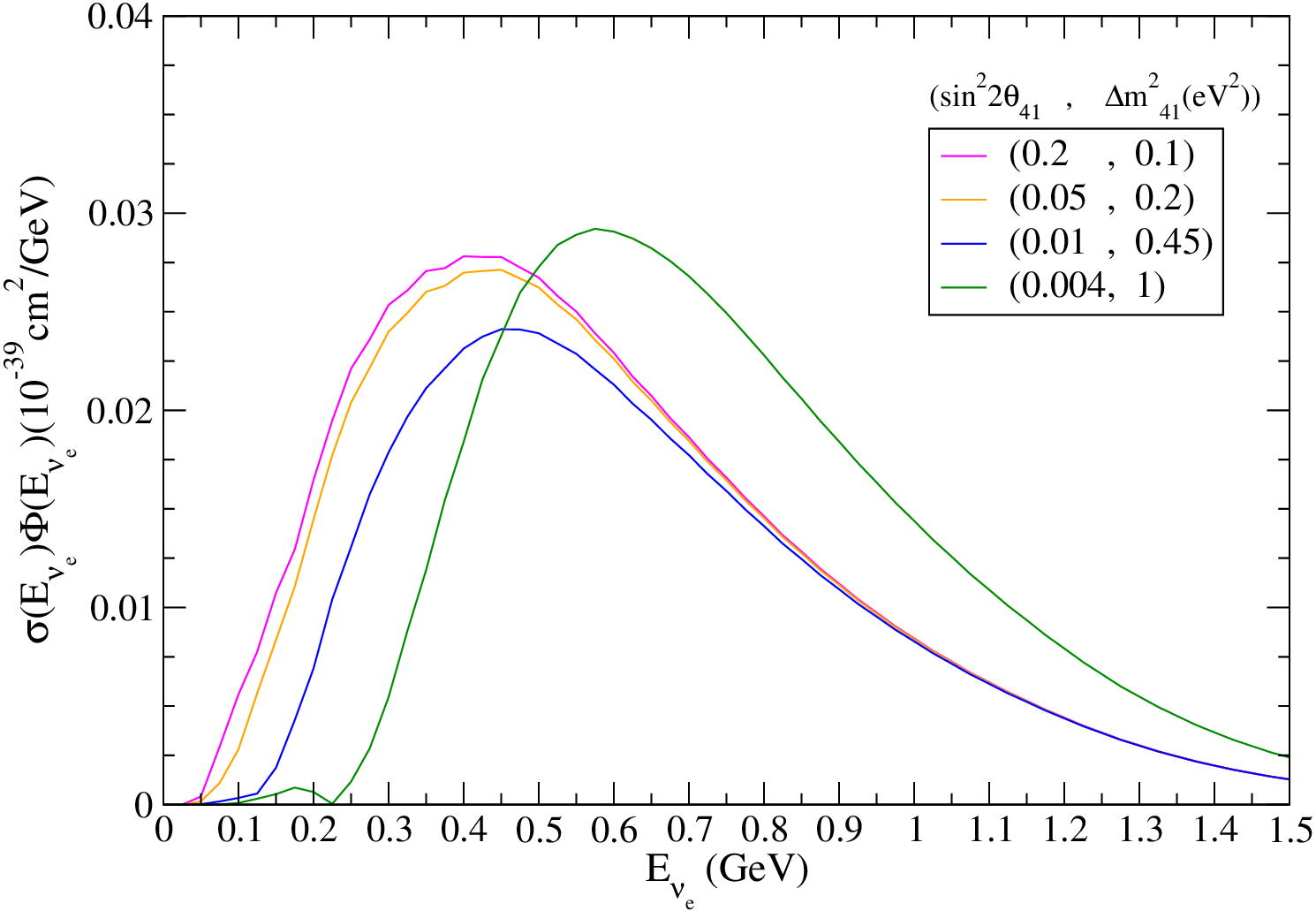}
\caption{(Color online) The electron distribution events $\sigma_{\nu_e}(E_{\nu_e}) \Phi_{\nu_e}(E_{\nu_e})$ corresponding 
to the fluxes of Fig. \ref{fig_art_minib_flux_osc}.
}
\label{fig_art_minib_sig_flux}
\end{center}
\end{figure}
  The theoretical 
 distributions of the electron events, the product of the ``quasielastic'' cross section by the oscillation electron 
neutrino energy distributions  are displayed in Fig. \ref{fig_art_minib_sig_flux}    
  for the same oscillation parameters as Fig. \ref{fig_art_minib_flux_osc}. 
  We now perform the smearing procedure according to the Eq. (\ref{rho_enubar}). 
\begin{figure}
\begin{center}
\includegraphics[width=12cm,height=8cm]{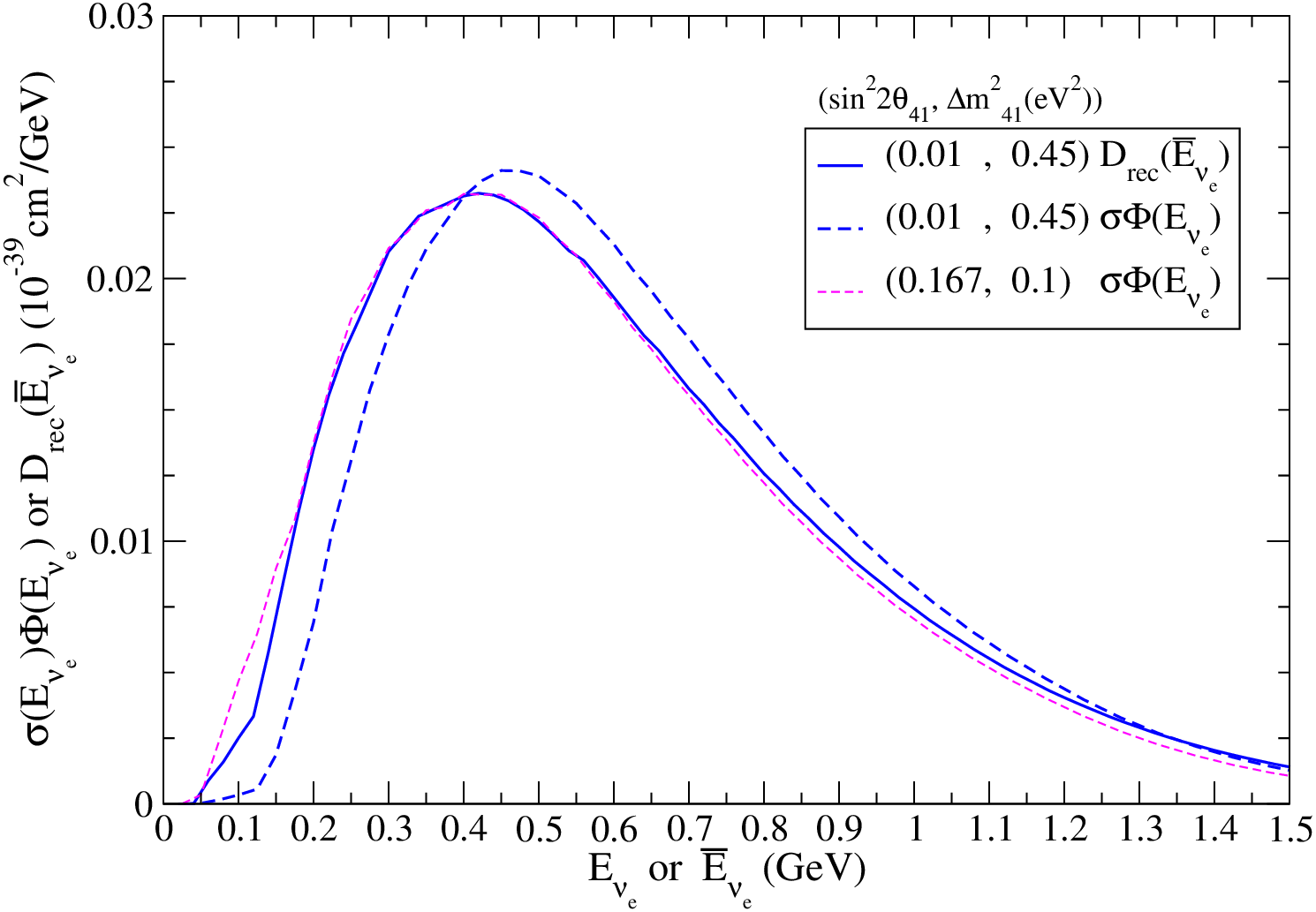}
\caption{(Color online) Effect of the smearing procedure for a mass parameter $ \Delta m_{41}^2=0.45$ eV$^2 $ and comparison with the effect of the lowering, in the unsmeared case, of the mass to the value $ \Delta m_{41}^2$= 0.1 eV$^2$.
}
\label{fig_art_minib_045}
\end{center}
\vspace*{+0.8cm}
\end{figure}
Figure \ref{fig_art_minib_045}  illustrates its effect for a mass parameter  $ \Delta m_{41}^2=0.45$ eV$^2 $. The smeared curve is shifted at lower energies, 
the displacement of the peak is appreciable, $\simeq 100$ MeV. The smearing effect increases the strength at low energies and decreases it at high ones. This goes in the right direction, as too much strength is observed at low energies and not enough at large ones. It is important to stress that the smearing effect is here equivalent and amounts to a lowering of the mass parameter. The smeared curve of Fig. \ref{fig_art_minib_045} can be reproduced in the unsmeared case 
with a value of the mass $  \Delta m_{41}^2=0.1$ eV$^2 $ as is illustrated in Fig. \ref{fig_art_minib_045}. 
We have here adjusted the associated values of $\sin^2(2\theta_{41})$ to reach  similar magnitudes. 
The equivalence is here complete. \textit{This means that, taking into account the smearing, a large mass value  $ \Delta m_{41}^2=0.45$  eV $^2 $ allows the same quality of  fit of the data than is obtained in the unsmeared case with a much smaller mass   $ \Delta m_{41}^2=0.1$  eV $^2 $}. 
Obviously there is an important gain for the compatibility with the existing constraints since the angle  $\theta_{41}$ can be smaller with a larger mass value. 
\begin{figure}
\begin{center}
\includegraphics[width=12cm,height=8cm]{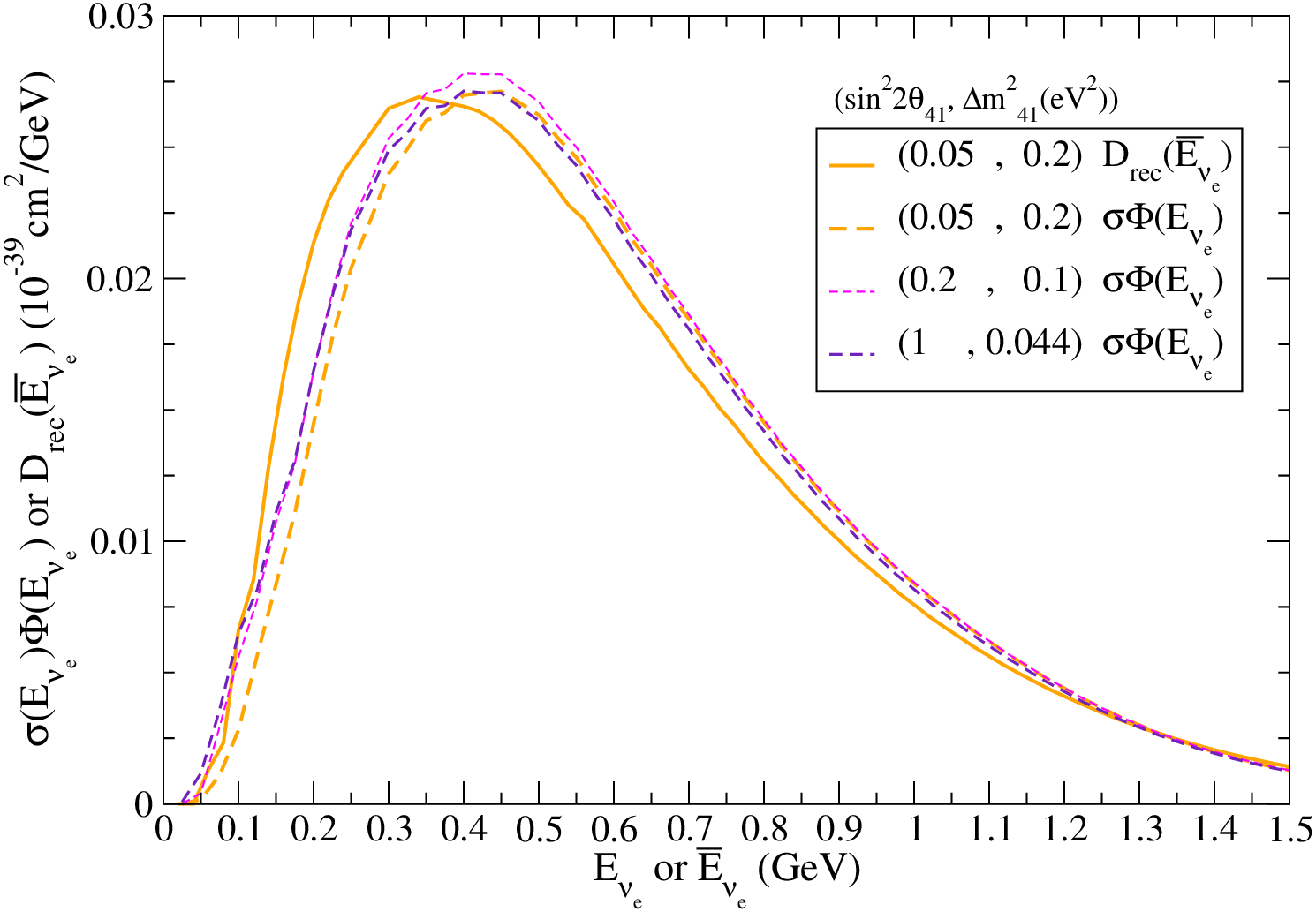}
\caption{(Color online) Effect of the smearing procedure for a mass parameter  $ \Delta m_{41}^2=0.2$  eV$^2$ and comparison with the lowering of the mass in the unsmeared case.
}
\label{fig_art_minib_02}
\end{center}
\end{figure}
  The cases of smaller mass values,  such as $ \Delta m_{41}^2=0.2$ eV$^2$, is also interesting. The corresponding smeared and unsmeared curves  
    are displayed in Fig. \ref{fig_art_minib_02} showing the low energy shift, the  peak position is lowered by $\simeq80$ MeV. 
What is interesting is that it is impossible to reproduce, in the region of interest $E_\nu > 0.2$ GeV,  the smeared curve with an unsmeared one even taking a very small mass. The reason is that, as mentioned before, there is a limiting energy shape of the unsmeared curve, as is apparent in Fig. \ref{fig_art_minib_02}, when the mass value is sufficiently small to reach the 
$1/E_{\nu_e}^2$ behavior for the energy dependence of the oscillated beam.
The depopulation by the smearing of the energy region above 400 MeV at the benefit of the region below, even if it is not spectacular, and beyond what is achievable by a lowering of the value of the oscillation mass, is susceptible to improve the $\chi^2$ value of the best fit. 
The influence of the smearing effect on the data, which goes in the right direction to account for the low energy anomaly, is the main message of our study of the MiniBooNE data and it should have an influence in the analysis of the data. 

This being said, there is an additional problem which naturally comes  to mind:  
the MiniBooNE muon neutrino beam contains a background of electron neutrinos which also undergo quasielastic events. 
These have  to be evaluated and subtracted from the observed  total number of electron events in order to extract the oscillation ones. 
These background electron events should be subject to the same reconstruction effects than the ones of the oscillated beam. 
This problem will be discussed in the following under the assumption that this background is evaluated in the  same way as the oscillation events, 
namely from the knowledge of the electron neutrino background energy distribution  and the ``quasielastic'' $\nu_e$ cross section. 
We are  aware however that this treatment does not apply as such to the MiniBooNE results as the background evaluated 
in the actual analysis has been the subject of a more detailed treatment  than the one sketched here, with the information from the muon events. 
However our discussion may focus the attention on some of the problems which can occur in the analysis.

\subsubsection{Inclusion of the MiniBooNE $\nu_e$ background}
\begin{figure}
\begin{center}
\includegraphics[width=12cm,height=8cm]{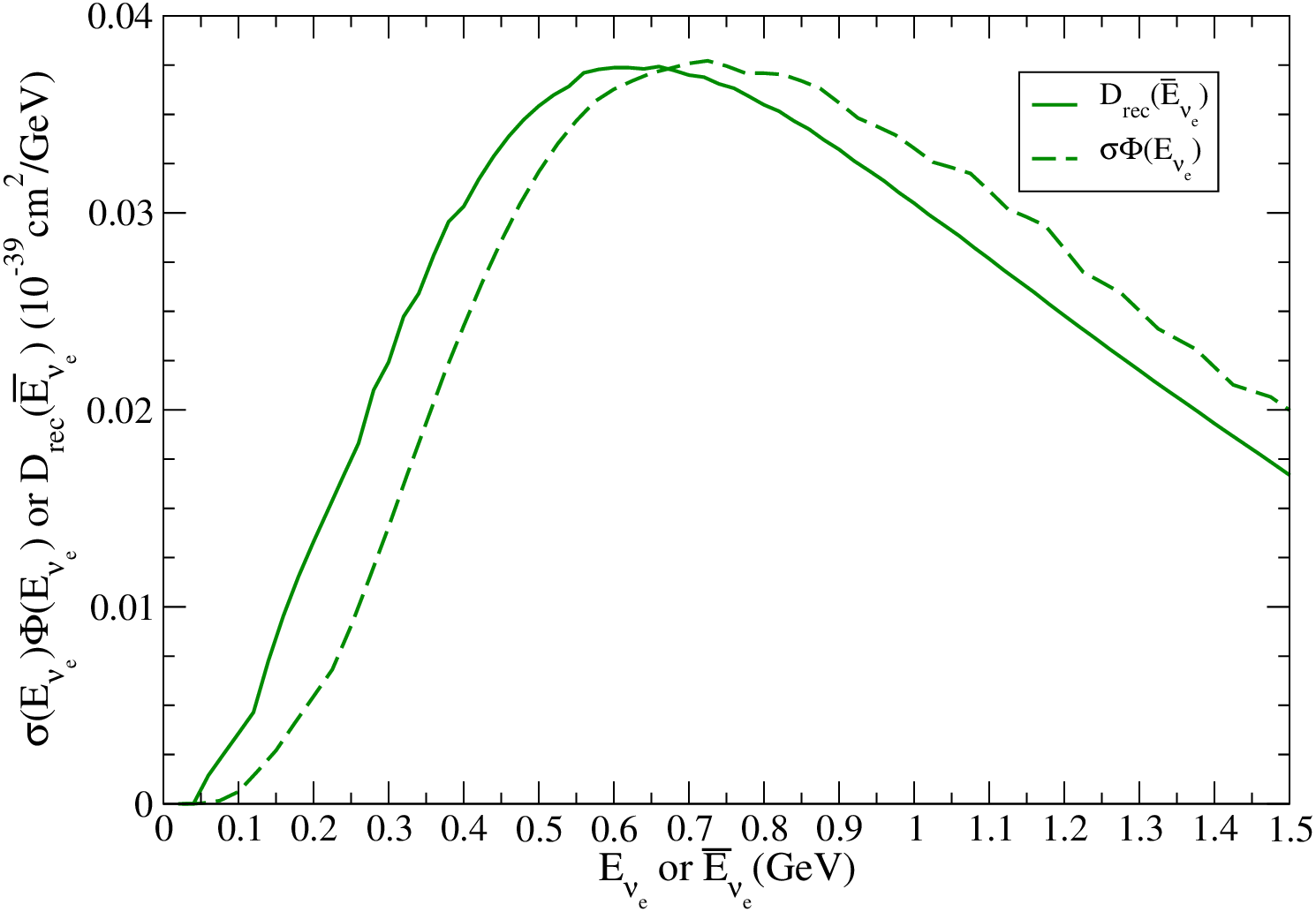}
\caption{(Color online) MiniBooNE electron events distribution for $\nu_e$ background in the unsmeared and smeared case.
}
\label{fig_art_minib_erec_siflu_nuebckgr}
\end{center}
\end{figure}
 
 We want to investigate the correction introduced by the transformation from real to reconstructed energy in the background 
 electron events distribution. In our previous work we have stressed the importance of the energy distribution in the reconstruction procedure. For the oscillation distribution,  depending on the oscillation parameter, it is rather narrow. While instead the   $\nu_e$ background  has a broad distribution.  
  As this distribution is different from the oscillation one, the reconstructed energy correction should be treated separately for the background and for the signal. We proceed for the background as for the signal. We start from the theoretical distribution for the $\nu_e$ background, 
$\sigma(E_{\nu_e}) \Phi_{background}(E_{\nu_e})$. We have scaled here the background flux given in Ref. \cite{AguilarArevalo:2008yp} 
by the same factor as the muon one, keeping in this way the right relative proportion between the background $\nu_e$ and the $\nu_\mu$ beams. 
We transform the $\sigma(E_{\nu_e}) \Phi_{background}(E_{\nu_e})$ in order to express it in terms of reconstructed energy. 
The outcome is shown in Fig. \ref{fig_art_minib_erec_siflu_nuebckgr} together with the theoretical distribution. 
As for the signal the smearing procedure increases the strength in the low energy region and reduces it beyond the peak position. This is essentially an effect of the np-nh piece, which is largely increased at small energies by the reconstruction procedure.
 We can therefore conclude that, in this description where the distribution of electron events from the $\nu_e$ background is evaluated theoretically as a product of the  $\nu_e$  cross section and of the background flux, which means that the reconstructed energy corrections for the background are ignored , the electron events 
background is underestimated for low reconstructed neutrino energies $E_{\nu_e}$ $\lesssim 0.6$ GeV and overestimated for larger ones. 
 Accordingly, in this picture,  the oscillation signal excess obtained by subtraction of this background from the total signal is overestimated in the low energy region and underestimated in the high energy one, which would be of great interest. However this conclusion is entirely linked to the assumption  made on the way in which the evaluation of the electron background is performed. Moreover, as this evaluation needs the absolute values, 
the cross section introduced is an essential ingredient. Here we have introduced the RPA cross section which introduces a quenching at low energies. It is partly compensated by the np-nh component, but the value of the cross section in the low energy domain remains an open question.
\begin{figure}
\begin{center}
\includegraphics[width=12cm,height=8cm]{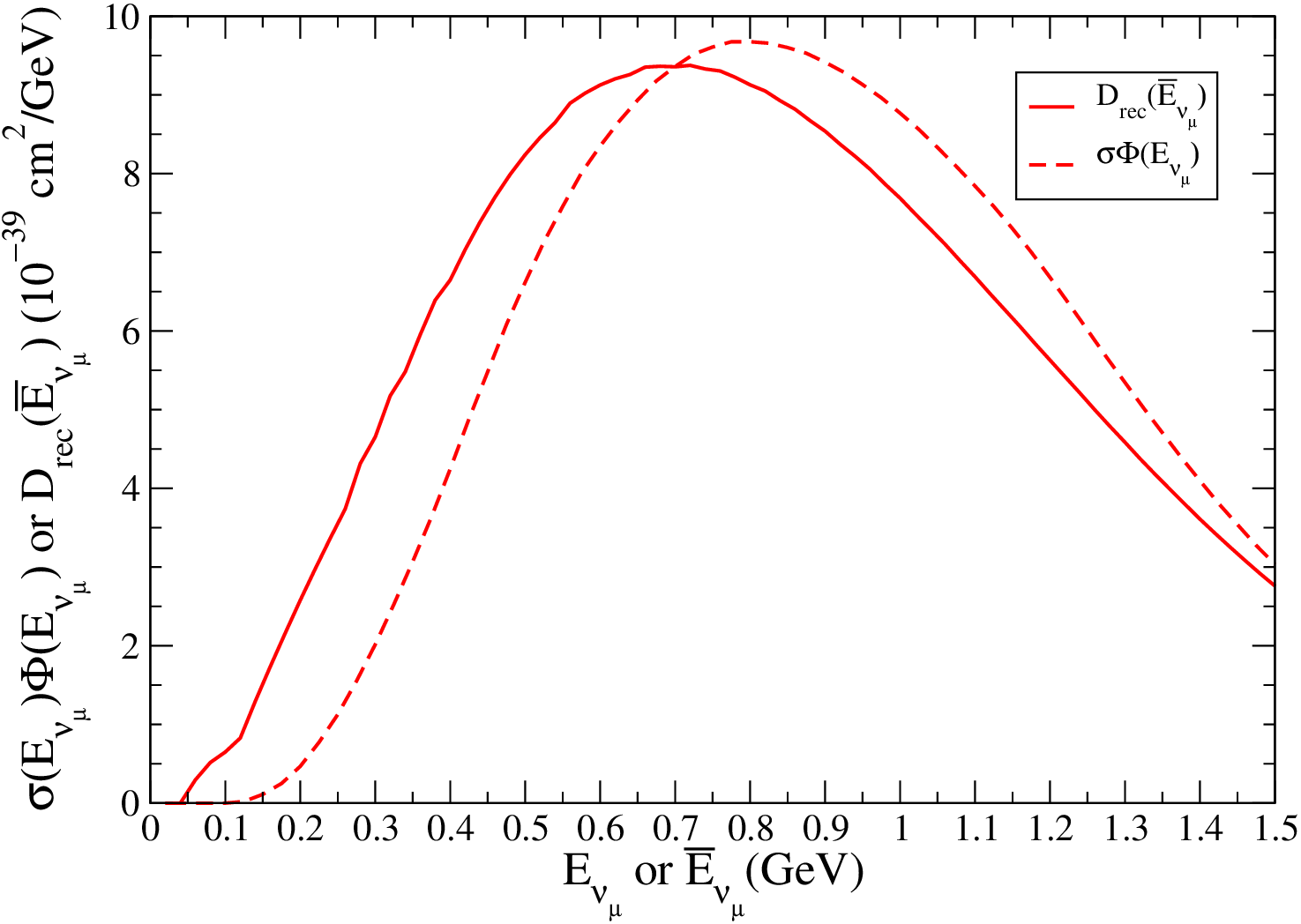}
\caption{(Color online) MiniBooNE muon events distribution in the unsmeared and smeared case.
}
\label{fig_art_minib_erec_siflu_mu.eps}
\end{center}
\end{figure}

 In addition, as already mentioned, the actual evaluation of the electron background relies in fact on a comparison with the muonic events, 
which also undergo the reconstruction correction. We give for completeness the reconstructed energy distribution also for the muonic events. 
The $\nu_\mu$ beam energy distribution is wider than the background  $\nu_e$ one. The effect of the smearing is then even more pronounced, particularly in the low energy region where the distribution is appreciably enhanced, as is illustrated in the Fig. \ref{fig_art_minib_erec_siflu_mu.eps}. 

\subsubsection{Effective cross sections}
\begin{figure}
\begin{center}
\includegraphics[width=12cm,height=8cm]{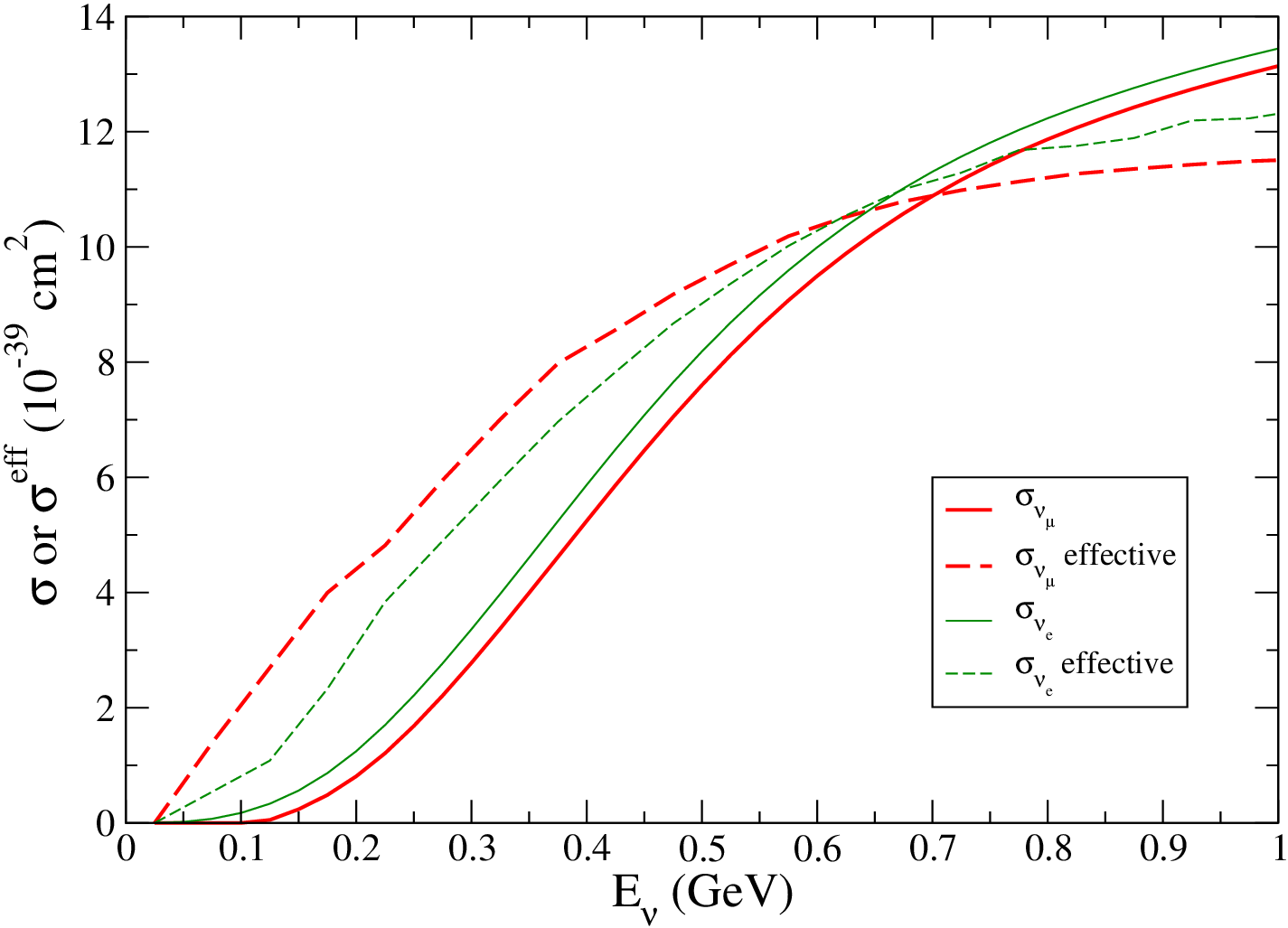}
\caption{(Color online) Free (unsmeared case) and effective (smeared case) $\nu$-$^{12}$C cross section per neutron for the MiniBooNE muonic events 
and for the background electron ones.
}
\label{fig_effective_sigma_muon_electron.eps}
\end{center}
\end{figure}
\begin{figure}
\begin{center}
\includegraphics[width=12cm,height=5.5cm]{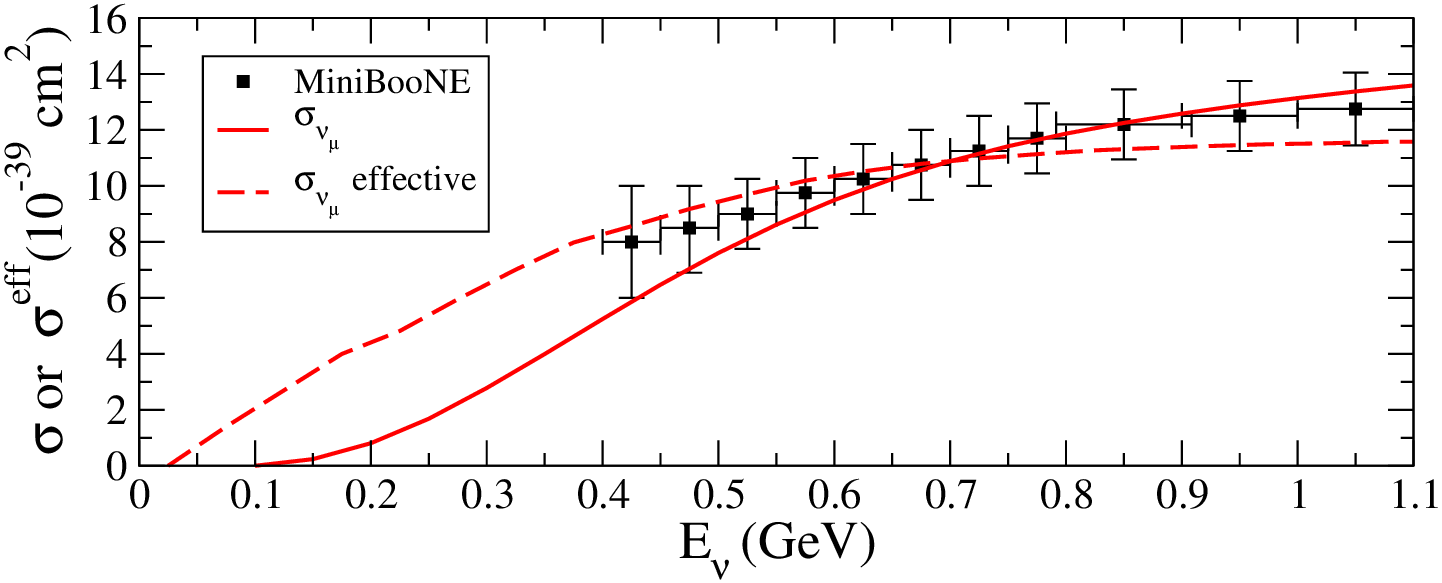}
\caption{(Color online) Free (unsmeared case) and effective (smeared case) $\nu$-$^{12}$C cross section per neutron for the MiniBooNE muonic events. 
The experimental MiniBooNE result \cite{AguilarArevalo:2010zc} is also shown.
}
\label{fig_effective_sigma_muon.eps}
\end{center}
\end{figure}

All the smeared distributions $ D_{rec}(\overline {E_{\nu} })$
can be expressed in terms of an effective ``quasielastic'' cross section $ \sigma_\nu^{eff} (\overline {E_{\nu} })$ 
(which also includes the multinucleon emission) 
with the following definition, which applies to muons as well as to electrons: 
$ D_{rec}(\overline {E_{\nu} }) = \sigma_\nu^{eff} (\overline {E_{\nu} }) \Phi(\overline {E_{\nu} })$.
 We can now suppress the specification $\overline {E_{\nu} }$ for the variable and write instead $E_{\nu}$. 
From the evaluation of  the  smeared distribution $ D_{rec}(\overline {E_{\nu} })$ 
 one deduces an effective cross section, which is different from the free one.  
This is the cross section  that should be associated to the flux distribution in a theoretical evaluation to be directly compared to the data when these are expressed with the reconstructed energy. Its use incorporates the reconstruction correction.  
We stress once more that this effective cross section is not universal but it  depends on the particular beam energy distribution. Therefore on the calculational side there is no gain with respect to the previous evaluation, as this cross section differs for each energy distribution.
 Applied to the electron background, the effective cross section is different than the muonic one because on the one hand the lepton produced (e) is lighter and on the other hand the energy distribution of which it depends is also different.  
Without reconstruction, the electron cross section surpasses somewhat the muonic one, particularly in the threshold region.
For the effective cross section the trend is reversed, the muonic one is larger than the background electron one, up to $E_{\nu} \simeq$ 0.6 GeV.
 This difference of trends which emerges in this analysis shows some potential problems which may arise in the evaluation of the electron background from the $\nu_e$ contamination of the beam. In Fig. \ref{fig_effective_sigma_muon.eps} we have also shown the MiniBooNE data points \cite{AguilarArevalo:2010zc} 
for the muonic cross section 
which is expressed with the reconstructed energy. The points should then be compared to the effective cross section for muons. 
As compared to our first work \cite{Martini:2009uj} where the reconstruction correction was ignored the agreement remains satisfactory as expected, 
since it reflects the detailed agreement that we have \cite{Martini:2011wp} for the double differential cross section .
   
\subsubsection{Antineutrinos}
\begin{figure}
\begin{center}
\includegraphics[width=12cm,height=8cm]{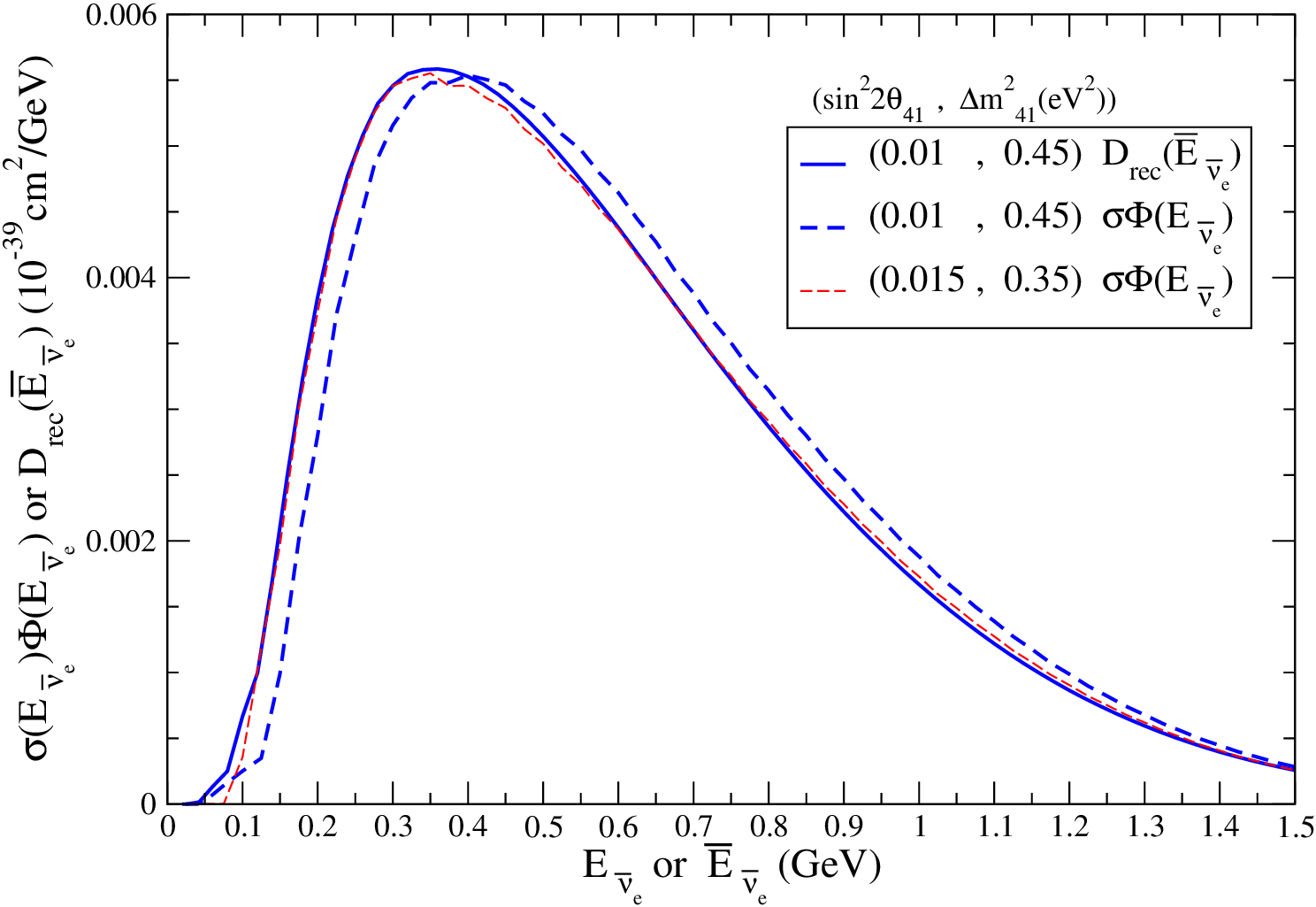}
\caption{(Color online) Effect of the smearing procedure for antineutrino for a mass parameter  $ \Delta m_{41}^2=0.45$  eV$^2 $ and comparison with the lowering of the mass in the unsmeared case to the value $ \Delta m_{41}^2$= 0.35 eV$^2$.
}
\label{fig_art_minib_erec_sigmaflux_anu_Dm_045_Dmfit}
\end{center}
\end{figure}

Similar effects, although  less pronounced, are present for antineutrino. 
They are not marked as in the neutrino case as, in our description, the np-nh events affect only the spin isospin response and are thus less influential 
in the reconstruction problem. For instance the smeared oscillation curve for a mass  parameter $ \Delta m_{41}^2=0.45$  eV$^2 $  is equivalent in shape to an unsmeared one with 
$ \Delta m_{41}^2$=0.35 eV$^2$, as shown in Fig. \ref{fig_art_minib_erec_sigmaflux_anu_Dm_045_Dmfit}. 
The gain is less than for neutrinos. But this awaits the test from the measurements of the antineutrino cross sections.

\section{Conclusion}

We have discussed the application of our reconstruction method to several oscillation experiments. 
Starting from a theoretical energy distribution, 
product of the neutrino cross section by the neutrino energy distribution of the beam, 
we perform the reconstructed energy corrections. Their influence is such that the events tend to escape from the region 
of high fluxes with a tendency to concentrate at lower energies. 
The smeared distribution effect depends on the particular shape of the neutrino energy distribution, the correction being more pronounced for broad distributions. 
We apply our procedure to the three distributions measured in T2K: muonic distributions in the close and far detector and electron distribution in the far detector. 
The effects are such that an analysis which takes into account the smearing effect 
is likely to lead to some increase of the oscillation mass value. 

We have also discussed the MiniBooNE results where the oscillations, if they exist, 
imply sterile neutrinos with a much larger mass parameter, in the eV range. 
The accumulation of electron events observed in the low energy sector favors relatively low values of this mass term which imply large mixing angles, 
not compatible with existing constraints. 
We have shown that this contradiction can be in part solved by the inclusion of the smearing effects. 
We have also pointed out some possible problems 
which may occur in the evaluation of the 
background from the $\nu_e$ contamination of the $\nu_\mu$ beam. In the antineutrino case where we predict similar effects but not as pronounced, 
the elucidation of the difference with neutrinos, which is important for the CP violation determination, will come from the measurement 
in progress of the double differential cross sections. In all instances the introduction 
of the smearing effect should improve the fit to the neutrino MiniBooNE data and its compatibility with the constraints from other sources. 
\\

Acknowledgments
\\

We are grateful to Carlo Giunti and Raymond Stora for interesting discussions.
We also thank Geralyn Zeller for useful information on the MiniBooNE data analysis. 
One of us (M.M.) acknowledges the hospitality of the CERN theory division where part of this work was done. 
This work was partially supported by the Communaut\'e Fran\c caise de Belgique (Actions de Recherche Concert\'ees).

\end{document}